\documentclass[preprint,onecolumn,showpacs,amsmath,amssymb,aps]{revtex4}

\usepackage{graphicx}
\usepackage{dcolumn}
\usepackage{bm}

\usepackage{amsmath}
\usepackage{amssymb}
\usepackage{url}
\usepackage{subfigure}
\usepackage{algorithm}
\usepackage{algorithmic}
\usepackage{enumerate}
\usepackage{slashbox}

\begin{document}

\preprint{APS/123-QED}

\title{Overlapping Community Detection in Complex Networks using Symmetric Binary Matrix Factorization}

\author{Zhong-Yuan Zhang}
\email{zhyuanzh@gmail.com}
\affiliation{School of Statistics, Central University of Finance and Economics}
\author{Yong Wang}%
\affiliation{Academy of Mathematics and Systems Science, Chinese Academy of Sciences}%
\author{Yong-Yeol Ahn}%
\email{yyahn@indiana.edu}
\affiliation{School of Informatics and Computing,
Indiana University Bloomington}%
\date{\today}

\begin{abstract}
Discovering overlapping community structures is a crucial step to understanding the structure and dynamics of many networks. In this paper we develop a \emph{symmetric binary matrix factorization} model (SBMF) to identify overlapping communities. Our model allows us not only to  assign community memberships explicitly to nodes, but also to distinguish outliers from overlapping nodes.
In addition, we propose a modified {\it partition density} to evaluate the quality of community structures. We use this to determine the most appropriate number of communities. We evaluate our methods using both synthetic benchmarks and real world networks, demonstrating the effectiveness of our approach.
\end{abstract}

\pacs{Valid PACS appear here}
\maketitle

\section{\label{Introduction}Introduction}
Many complex networks possess community structure \cite{Girvan02}.
Intuitively speaking, a community is a set of nodes that are densely interconnected but loosely connected with the rest of the network. Nodes often belong to more than one community, causing communities to overlap with each other \cite{overlap,link}. For example, a person in a social network belongs to many groups, such as a family, a company, and a baseball club; many proteins have more than one function and belong to multiple functional communities \cite{protein}. The overlap cannot be detected by the traditional hard-partitioning methods that assign each node to a single community \cite{ZHANG}.

Because of its abundance,  overlapping community structure has become a hot research topic and many detection methods have been introduced. They can be classified into two categories based on their outputs \cite{fuzzy}: fuzzy overlapping community detection and non-fuzzy overlapping community detection. Fuzzy overlapping community detection methods estimate the strength of memberships, while not being able to provide clear node membership to each community \cite{bridgeness,nmf1,HJL}. By contrast, non-fuzzy overlapping community detection methods give crisp partitions, allowing each node to have multiple community labels \cite{gnmi,link}. These methods do not provide any information about the strength of the nodes' membership to each community. In short, each approach has complementary benefits and drawbacks, raising a question: Can we combine the advantages of both kinds of approaches?

In this paper, we propose a \emph{Symmetric Binary Matrix Factorization} (SBMF) model to combine the best of both. It is motivated by both Symmetric Nonnegative Matrix Factorization (SNMF) \cite{nmf99,nmf01} and Binary Matrix Factorization (BMF) \cite{BMF, BMF2}.
The SBMF enables us not only to  identify the community structures explicitly, but also to analyze the strength of membership based on the corresponding results from SNMF, providing a comprehensive picture exploiting both SNMF and SBMF. Furthermore, the model can distinguish the outliers, which do not belong to any communities, from the overlapping ones.

An essential step in community detection is defining the quality of the detected community structure. One of the most widely used quantities is the modularity function $Q$ \cite{Girvan04,Newman06}. However, the modularity function has several drawbacks. The most well-known problem is the resolution limit, such that $Q$ cannot capture small communities, however strongly clustered they are \cite{resolution}. Several other issues of $Q$ have been identified including bottleneck-dependence \cite{bottlenecks},
misidentification of communities \cite{ZHANG2}, and degeneracy problem \cite{performance}. 
Recently, \emph{partition density}, a new measure designed for evaluating highly overlapping communities, was proposed \cite{link}. In this paper we use a modified partition density to assess node-based communities instead of link communities.

In summary, the contributions of this paper are: (i) proposing a parameter-free, simple-to-implement overlapping community detection method, SBMF;
(ii) providing a method to infer an appropriate number of communities using modified partition density; and (iii) demonstrating the effectiveness of the proposed method by systematically conducting experiments on both the synthetic and the real-world networks.

The rest of the paper is organized as follows: Section \ref{BMF} introduces the symmetric binary matrix factorization. Section \ref{MS} presents the new partition density. Section \ref{toy} explains an illustrative example. Section \ref{results} shows the experimental results. Finally, Section \ref{conclusion} concludes.
\section{\label{BMF}Symmetric Binary Matrix Factorization}
\subsection{Motivation}
Let us begin from a matrix factorization approach for community detection \cite{bridgeness}:
\begin{equation}
\begin{array}{rl}\label{Eq:02}\vspace{2mm}
\min & \|X-UU^T\|_F^2\\\vspace{2mm}
s.t. & U\geq0,\\
     & \sum\limits_{j=1}^cU_{ij}=1, i = 1,2,\cdots,n.
\end{array}
\end{equation}
where $X$ is the adjacency matrix of the network of size $n\times n$, and $U$ is the community membership matrix of size $n\times c$. Note that in this paper, we set the diagonal elements of $X$ to $1$ in accordance with the assumption that the connected pairs of nodes are more similar to each other.

The standard nonnegative matrix factorization (NMF) tries to factorize a nonnegative matrix $X$ of size $n\times m$ into two nonnegative matrices, $U_1$ of size $n\times c$ and $U_2$ of size $m\times c$, such that $X\approx U_1U_2^T$ \cite{nmf99, nmf01}.
For the symmetric objective matrix $X$, NMF can be reduced to symmetric NMF (SNMF; see Eq. (\ref{Eq:02})).

 NMF is becoming one of the most popular and widely accepted models in unsupervised learning \cite{nmf99, gene, tensor}. Several NMF-based models have been applied to community detection. Thanks to the flexibility of the model, NMF is particularly suitable for the detection of  overlapping communities \cite{nmf1, nmf2, nmf3, nmf4}. The result $U$ can be interpreted as a cluster membership degree matrix, i.e., node $i$ belongs to a community $t$ with the strength $U_{it}$ (note that $\sum_jU_{ij}=1$), providing ``fuzzy'' overlapping communites. But, how to determine whether a node really belongs to a community or not? Does the strength $0.9$ mean that the node belongs to the community? What about the strength $0.5$?  It is often more useful to identify community structure explicitly and consider nodes as  full members of their communities \cite{link}. Another issue is that one cannot distinguish the outliers from the overlapping nodes based on the result $U$. For example, if a node's membership strength vector is $[1/c,1/c,\cdots,1/c]$, one cannot tell whether the node belongs to all of the communities with same strength or does not belong to any of them.

To address these problems, we introduce the SBMF model. The model can explicitly assign community memberships to nodes and can distinguish outliers from overlapping nodes.
\subsection{Model Formulation}
The SBMF can be defined as follows: given a symmetric binary matrix $A$ of size $n\times n$, we want to find a binary matrix $U$ of size $n\times c$ such that $A\approx UU^T.$  The objective matrix $A$ is the adjacency matrix of the network, and $U$ is the community membership indicator matrix: $U_{it}=1$ if node $i$ is in the community $t$, and $0$ if not. If a node $i$ belongs to multiple communities, then the sum of the corresponding row $i$ of $U$ will be larger than one ($\sum_jU_{ij}>1$). On the contrary, if the node $i$ is an outlier, the corresponding row $i$ will be zero ($\sum_jU_{ij}=0$).

We assume that there are a relatively small number of outliers, which do not belong to any communities in the network, and require that $U$ should have as few zero rows ($\sum_jU_{ij}=0$) as possible. We achieve this by adding a penalty term into the optimization model.  We use 1-norm \footnote{1-norm of a matrix $X$ is the largest column sum of $\mbox{abs}(X)$, where $\mbox{abs}(X)_{ij}=\mbox{abs}(X_{ij})$, and $\mbox{abs}(\cdot)$ is the absolute value.} instead of Frobenius-norm because it gives better numerical results.

In summary, SBMF can be formulated as the following constrained non-linear programming:
\begin{equation}
\begin{array}{rl}\label{Eq:04}\vspace{1mm}
\min\limits_{U} & \|A-UU^T\|_1+\sum\limits_i(1-\Theta(\sum\limits_jU_{ij}))\\
s.t. & U_{ij}^2-U_{ij}=0,\, i = 1,2,\ldots,n,\, j=1,2,\ldots,c,
\end{array}
\end{equation}
where $\Theta$ is the Heaviside step function: for some matrix $X$,
$$
\Theta(X)_{ij}:=\left\{\begin{array}{rl}\vspace{2mm}
                  1 & \mbox{if}\, X_{ij}>0;\\
                  0 & \mbox{if}\, X_{ij}\leqslant0.
                 \end{array}
                 \right.
$$
As one can see, all the elements of $U$ are variables that need to be
decided, which means that it is a large scale optimization problem. Furthermore, the function $\Theta$ is not continuous and the problem is non-smooth. Hence,
the standard optimization algorithms are not suitable. To overcome these difficulties, we firstly initialize $U$ by solving the NMF model (\ref{Eq:02}), then ``discretize'' it by solving the following simpler unconstrained non-linear programming, instead of (\ref{Eq:04}), to get an optimal approximation solution:
\begin{equation}
\begin{array}{rl}\label{Eq:01}\vspace{1mm}
\min\limits_{u} & \|A-\Theta(U-u)\Theta(U-u)^T\|_1+\\
 & \hspace{35mm}+\sum\limits_i(1-\sum\limits_j\Theta(U-u)_{ij}),
\end{array}
\end{equation}
where
$u$ is a scalar.

 To solve (\ref{Eq:01}), we fix $U$, and discretize the domain $\{u: 0\leqslant u\leqslant \max(U)\}$ to select $\hat{u}$ that minimizes the optimization problem (\ref{Eq:01}). Finally, we obtain the binary matrix $U$ as follows: $$U: =\Theta(U-\hat{u}).$$

To initialize $U$, we employ the algorithm of multiplicative update rules developed for SNMF, which is summarized in Algorithm \ref{Al:01}. We set the iteration number {\texttt iter} to 100 in this paper.

\begin{algorithm}[H]
\caption{Symmetric Nonnegative Matrix Factorization
 (Least Squares Error)}
\label{Al:01}
\begin{algorithmic}[1]
\REQUIRE $A,$ iter
\ENSURE $U$
\FOR{$t=1:\mbox{iter}$}\vspace{2mm}
\STATE
$
\displaystyle U_{ij}:=U_{ij}\frac{(AU)_{ij}}{(UU^{T}U)_{ij}}
$\vspace{2mm}
\STATE
$
\displaystyle{U_{ij}:=\frac{U_{ij}}{\sum_jU_{ij}}}\vspace{2mm}
$
\ENDFOR
\end{algorithmic}
\end{algorithm}

\section{\label{MS}Model Selection}
The problem of community detection is an unsupervised learning task, and the number of communities is unknown in real applications. Several methods have been developed to infer the number of communities. The most popular one uses the modularity function $Q$ \cite{Girvan04,Newman06}. Namely, choosing the number of communities $c$ at which the modularity function achieves the maximum. However, as explained above, the modularity function may lead to both the underestimation and overestimation of community numbers \cite{bad}. To overcome these limitations,
we modify partition density to better estimate the appropriate number of communities. Originally, the partition density was defined in terms of edges \cite{link}.

Formally, the standard partition density of the community $\alpha$ can be defined as \cite{link}:
$$
   D_\alpha = \frac{m_\alpha-(n_\alpha-1)}{n_\alpha(n_\alpha-1)/2-(n_\alpha-1)},
$$
where $n_\alpha$ and $m_\alpha$ are the number of nodes and the number of edges in the community $\alpha$, respectively.
Then the overall partition density of the network can be defined as the weighted sum of $D_\alpha,\, \alpha=1,2,\ldots,c$. The weight for each community was given by the number of edges in the community in the original formulation. Here we use the number of nodes as the weights because this formulation gives us better results. Thus, the partition density of the whole network is:
$$
   D = \frac{2}{N}\sum_{\alpha=1}^c n_\alpha\frac{m_\alpha-(n_\alpha-1)}{(n_\alpha-2)(n_\alpha-1)},
$$
where $N$ is the sum of the sizes of different communities and the number of outliers. $N$ may be larger than the number of nodes $n$ in the network, since the overlapping nodes are counted more than once.

Nodes can be assigned to multiple communities. However, in practice one can observe that the nodes with several community labels are not common due to limited energy, time, and resources. Hence, we add a penalty term into the denominator to control the nodes' activity degree, and the definition can thus be updated as follows:
$$
   D = \frac{2}{N}\sum_{\alpha=1}^c \frac{n_\alpha}{q_\alpha}\frac{m_\alpha-(n_\alpha-1)}{(n_\alpha-2)(n_\alpha-1)},
$$
where $q_\alpha=\max_{j\in \alpha}l_j$, $j\in\alpha$ means that the node $j$ belongs to the community $\alpha$, $l_j$ is the number of community labels that the node $j$ has.


Since the partition density $D$ only considers the local information in each community, it does not suffer from the resolution limit problem \cite{link}.
\section{An Illustrative Example}\label{toy}
\begin{figure*}
\subfigure{\includegraphics[height=60mm,width=140mm]{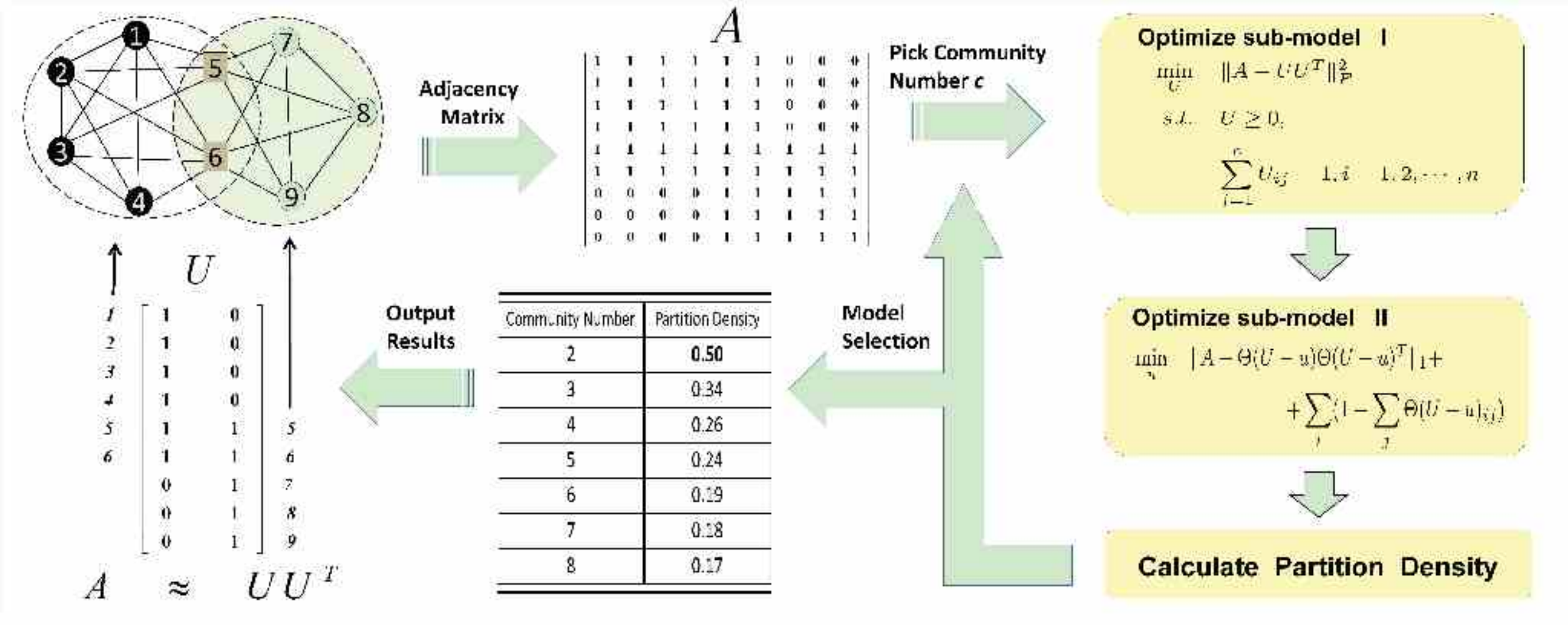}}
\caption{An illustrative example, with overlapping nodes $5$ and $6$, to show how the results of the proposed SBMF model can reveal the community structure in complex networks.}\label{Fig:02}
\end{figure*}

In this section we use a toy example to illustrate how the proposed method works. Figure \ref{Fig:02} shows a toy network with two communities, where Node $5$ and Node $6$ are bridges between them. First, we build the adjacency matrix $A$ and solve the SBMF model (\ref{Eq:04}) with different numbers of communities $c$. Then, we calculate the partition density for each $c$ to select the best one. Finally, the corresponding binary matrix $U$ is obtained. As one can observe, the two overlapping communities can be explicitly recovered from the output $U$ of SBMF model: 1 means that the node belongs to the corresponding community, and 0 means not.
\section{\label{results}Experimental Results}
In this section we test the effectiveness of our
method on both the synthetic and the real world networks.
\subsection{Datasets Description}
\begin{enumerate}[a)]
\item GN benchmark network \cite{Girvan02}: GN network has four equally-sized non-overlapping communities with 32 nodes each. On average each node has $Z_{in}$ edges connecting with the others in its own community and $Z_{out}$ edges connecting with the other three communities. As expected, the communities become less clear with increasing $Z_{out}$. Here $Z_{in}+Z_{out}$ is set to $16$.
\item LFR benchmark network \cite{LFR}: The GN networks do not capture most characteristics of real networks: size of the network, heterogeneous degree distribution, etc. And the LFR benchmark model was proposed to address it. In LFR benchmarks both the degree and the community size distributions obey power laws with exponents $\gamma$
and $\beta$. Each node has a fraction $1-\mu$ of its neighbors in its own community and a fraction $\mu$ in the other communities. Furthermore, nodes can be assigned to multiple communities.

In this paper we set the parameters of the LFR benchmark as follows: The number of nodes is $1000$, the maximum degree is 50, the exponent of the degree distribution $\gamma$ is 2 and that of the community size distribution $\beta$ is 1. We set the parameters for the non-overlapping LFR benchmark as follows: The average degree of the nodes is 20, and the range of the mixing parameter $\mu$ is from $0.1$ to $0.6$. For the overlapping LFR benchmark, we set the parameters as follows: The average degree of the nodes is from $15$ to $25$, the mixing parameter $\mu$ is $0.1$, the minimum community size is $20$, the maximum community size is $100$, and the fraction of the overlapping nodes is from $0.1$ to $0.5$.
\item Football \cite{Girvan02}: This dataset is the network of 115 American football teams. There are $613$ edges. The teams are represented as nodes, and the nodes are connected if there is a game between them.  The teams are divided into 12 conferences, where the teams usually play more games with others in the same conference, inducing community structure.
\item Political Books \footnote{http://www.orgnet.com/cases.html}: This dataset is the Amazon co-purchasing network with $105$ books about US politics. There are $441$ edges.
Nodes are books and edges represent co-purchasing of books by the same
buyers. 
\item Dolphins \cite{dolphins}: This dataset is the social communication network of 62 bottlenose dolphins that lived in Doubtful Sound, New Zealand. There are $159$ edges.
\item Jazz Bands \cite{jazz}: This dataset is the collaboration network of jazz bands. There are $198$ nodes representing the bands, and $2742$ edges connecting the bands if there is at least one musician in common.
\end{enumerate}
\subsection{Assessment Standards}
To evaluate the detection
performance on non-overlapping synthetic networks (GN \& non-overlapping LFR), we use the normalized mutual information (NMI) \cite{nmi}. The value can be formulated as follows:
$$\displaystyle
  NMI(M_1, M_2) = \frac{\sum\limits_{i=1}^{k}
     \sum\limits_{j=1}^{k}n_{ij}\log\displaystyle\frac{n_{ij}n}{n_i^{(1)}n_j^{(2)}}}
  {\sqrt{\left(\sum\limits_{i=1}^{k}n_i^{(1)}\log\displaystyle\frac{n_i^{(1)}}{n}\right)
  \left(\sum\limits_{j=1}^{k}n_j^{(2)}\log\displaystyle\frac{n_j^{(2)}}{n}\right)}},
$$
where $M_1$ and $M_2$ are the ground-truth community label and the computed community label, respectively; $k$ is the community number; $n$ is the number of nodes; $n_{ij}$ is the number of nodes in the ground-truth community $i$ that are assigned to the computed community $j$; $n_i^{(1)}$ is the number of nodes in the ground-truth community $i$;  $n_j^{(2)}$ is the number of nodes in the computed community $j$; and $\log$ here is the natural logarithm. The larger the NMI value, the better the community partition.

For overlapping LFR benchmarks, we use the generalized normalized mutual information {\cite{gnmi}}.
\begin{figure*}
\subfigure{\includegraphics[height=50mm,width=81mm]{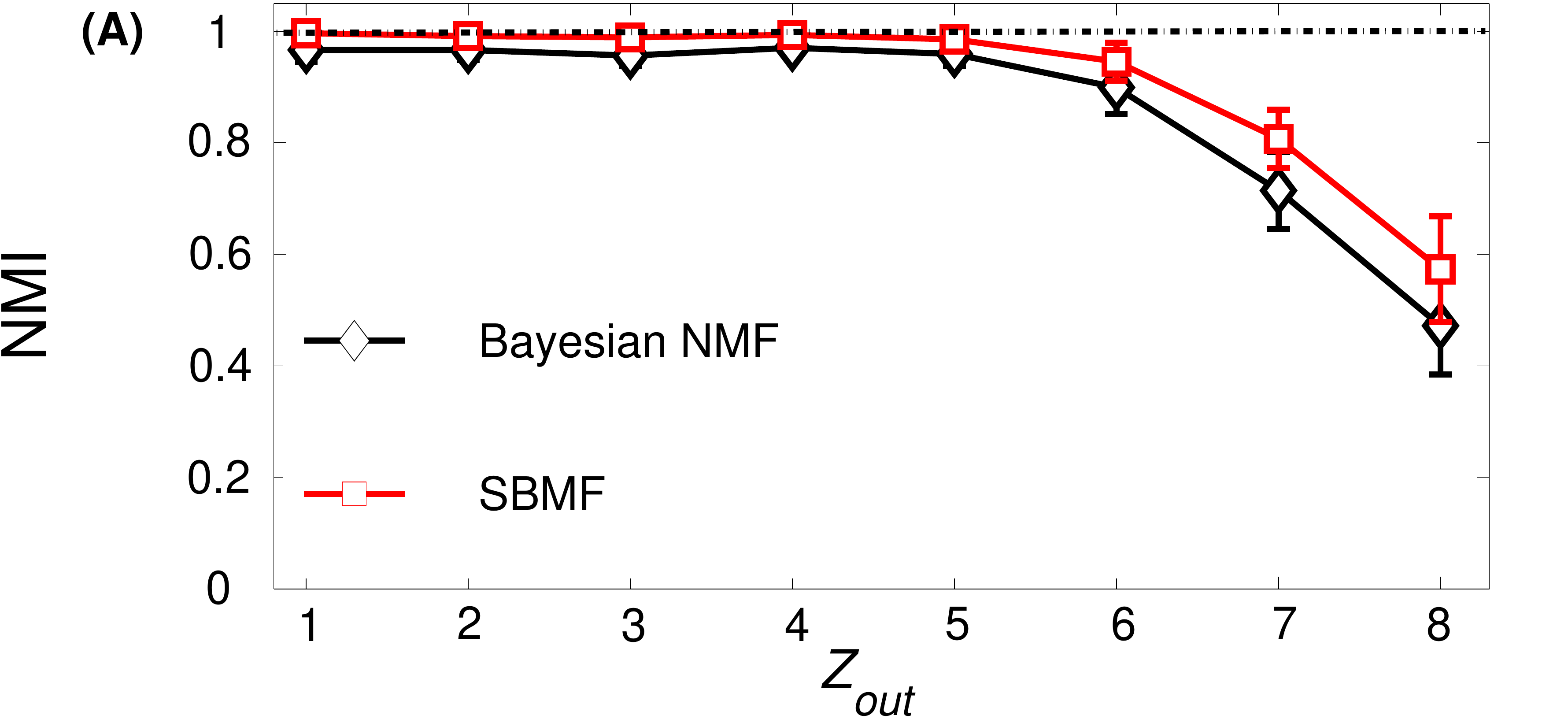}}
\subfigure{\includegraphics[height=50mm,width=81mm]{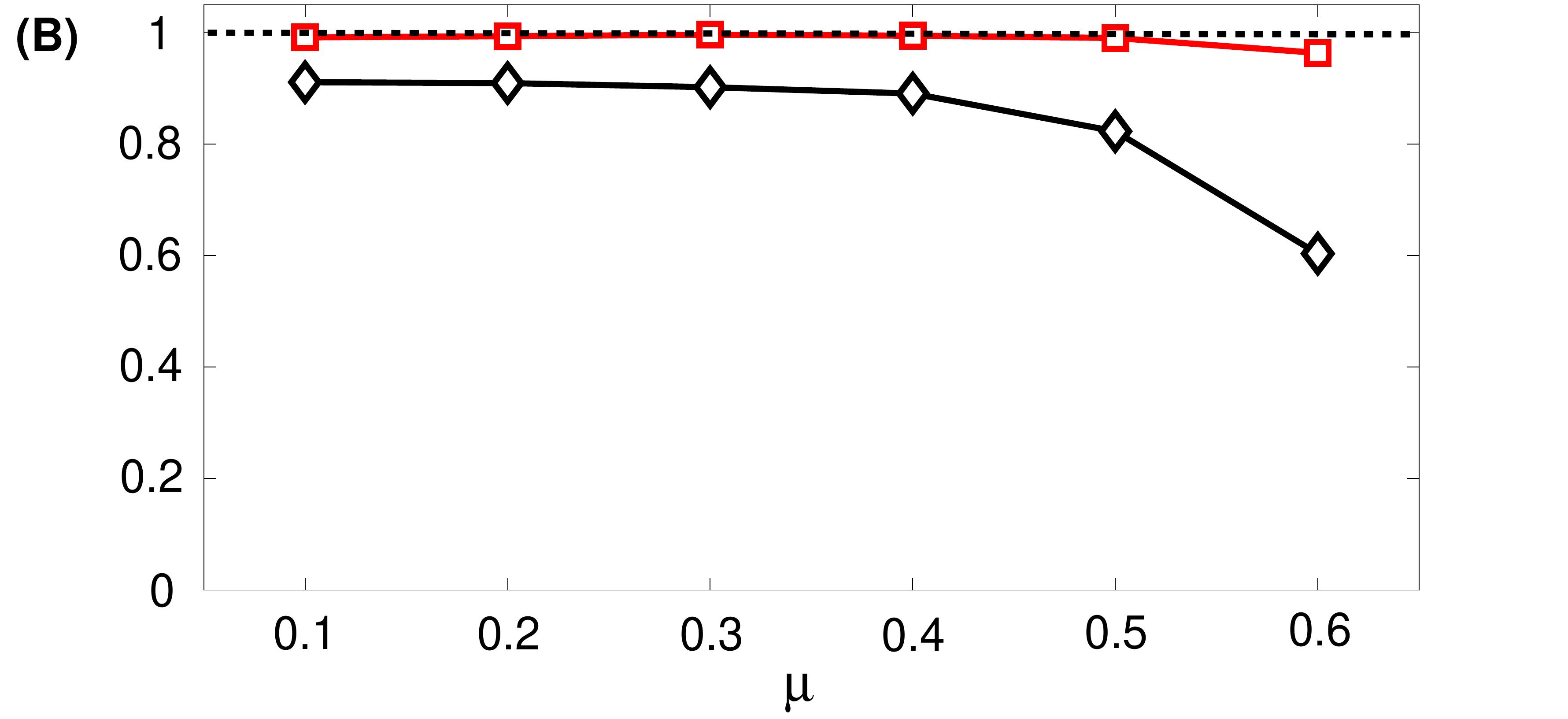}}
\caption{Averaged NMI with the standard deviation of Bayesian NMF and SBMF on (A) GN networks and (B) non-overlapping LFR networks.}\label{Fig:03}
\end{figure*}
\subsection{Experimental Results on Synthetic Networks}
\begin{figure}
 \centering
\includegraphics[height=47mm,width=83mm]{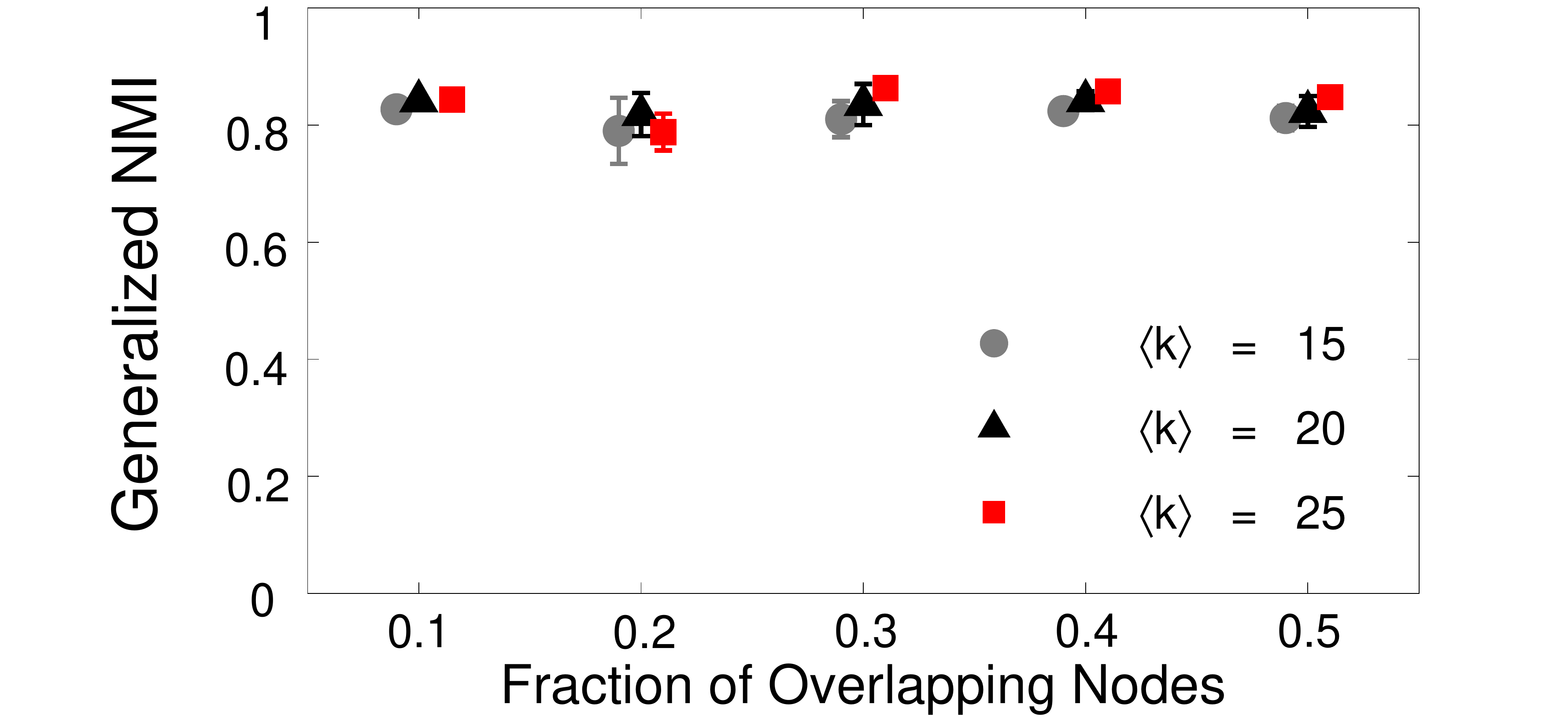}
\caption{Averaged generalized NMI with the standard deviation of SBMF on the overlapping LFR networks. ``$\langle k\rangle$'' means average degree.}\label{Fig:04}
\end{figure}

\begin{figure}
\centering
\includegraphics[height=50mm,width=90mm]{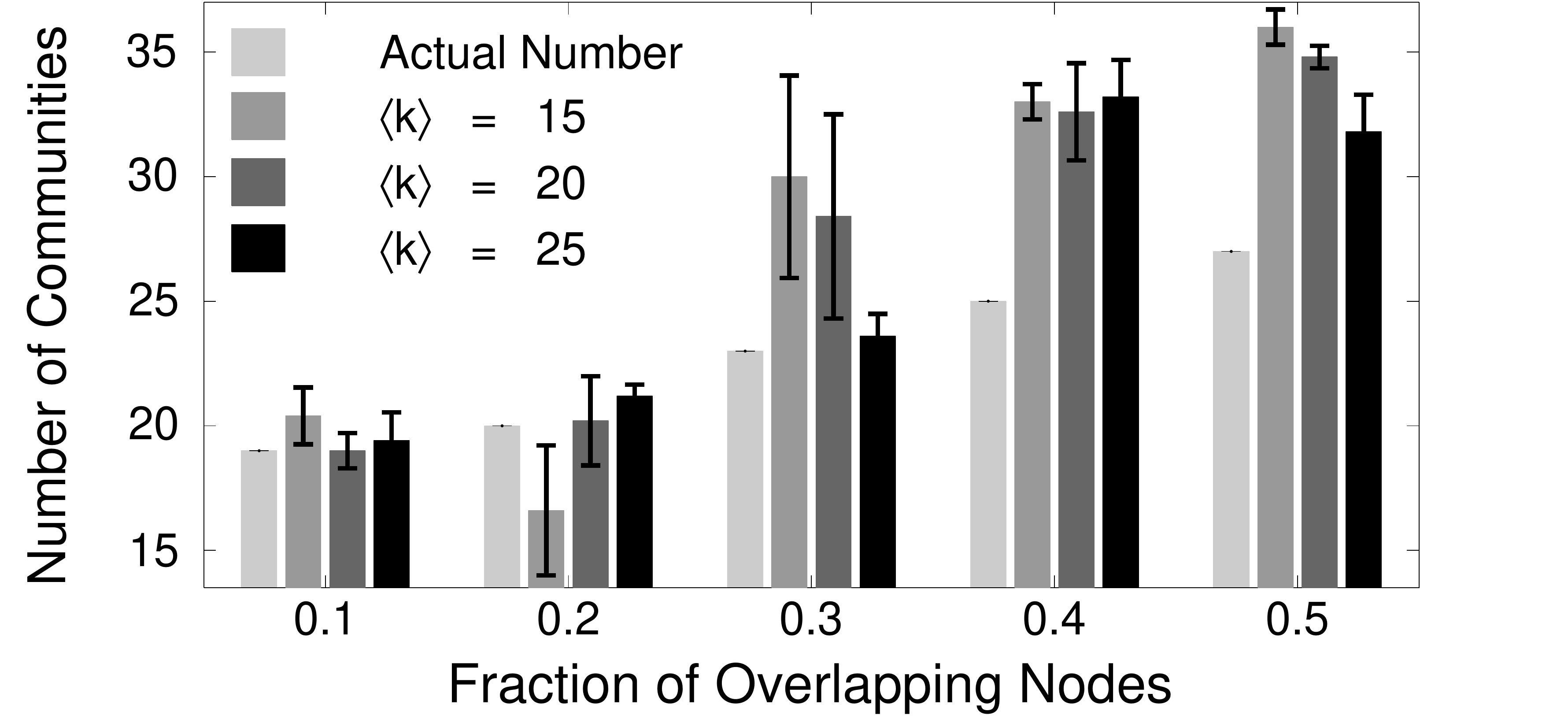}
\caption{Number of communities estimated by SBMF on overlapping LFR networks with the standard deviation errors. ``$\langle k\rangle$'' represents the average degree of nodes in the network, and ``Actual Number'' represents the actual number of communities in the corresponding LFR network.}\label{Fig:10}
\end{figure}

\begin{figure*}
\centering
\subfigure{\includegraphics[height=50mm,width=75mm]{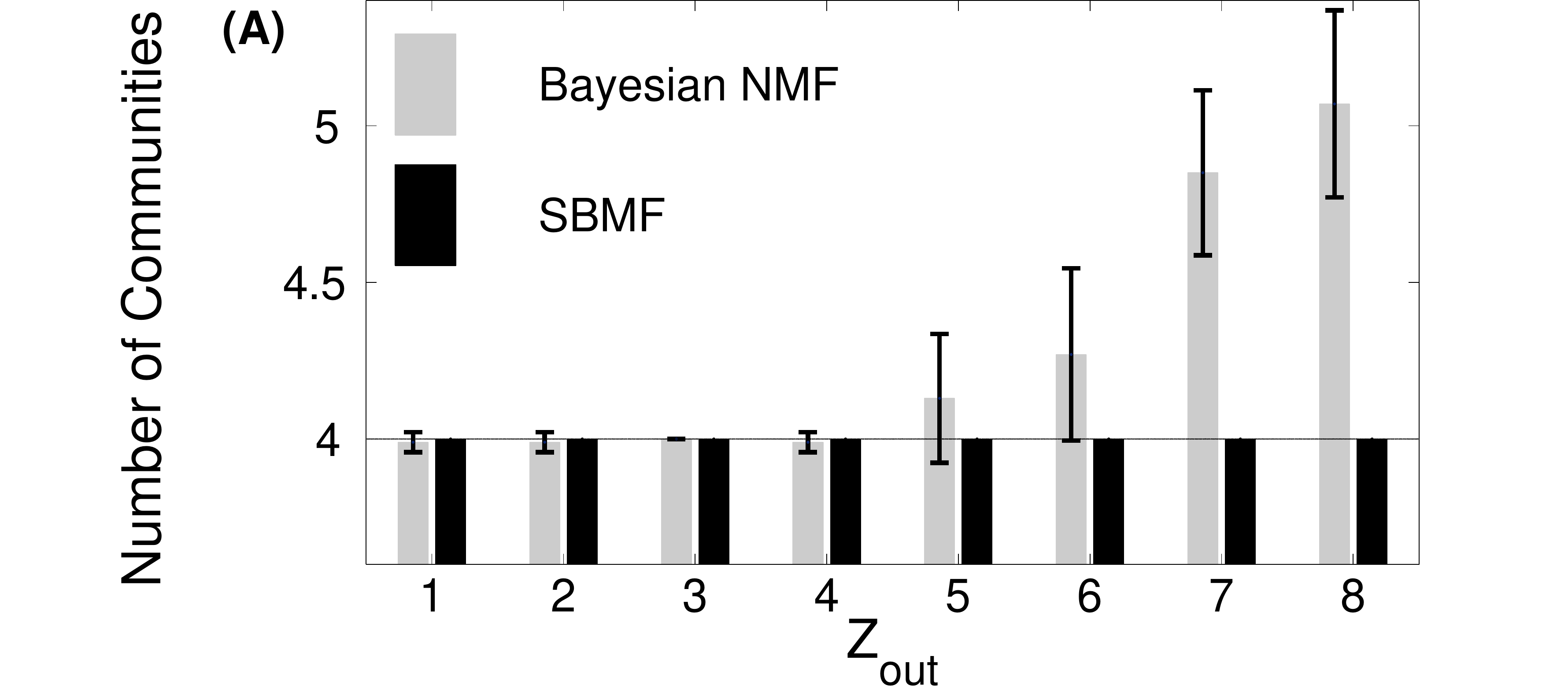}}
\subfigure{\includegraphics[height=50mm,width=75mm]{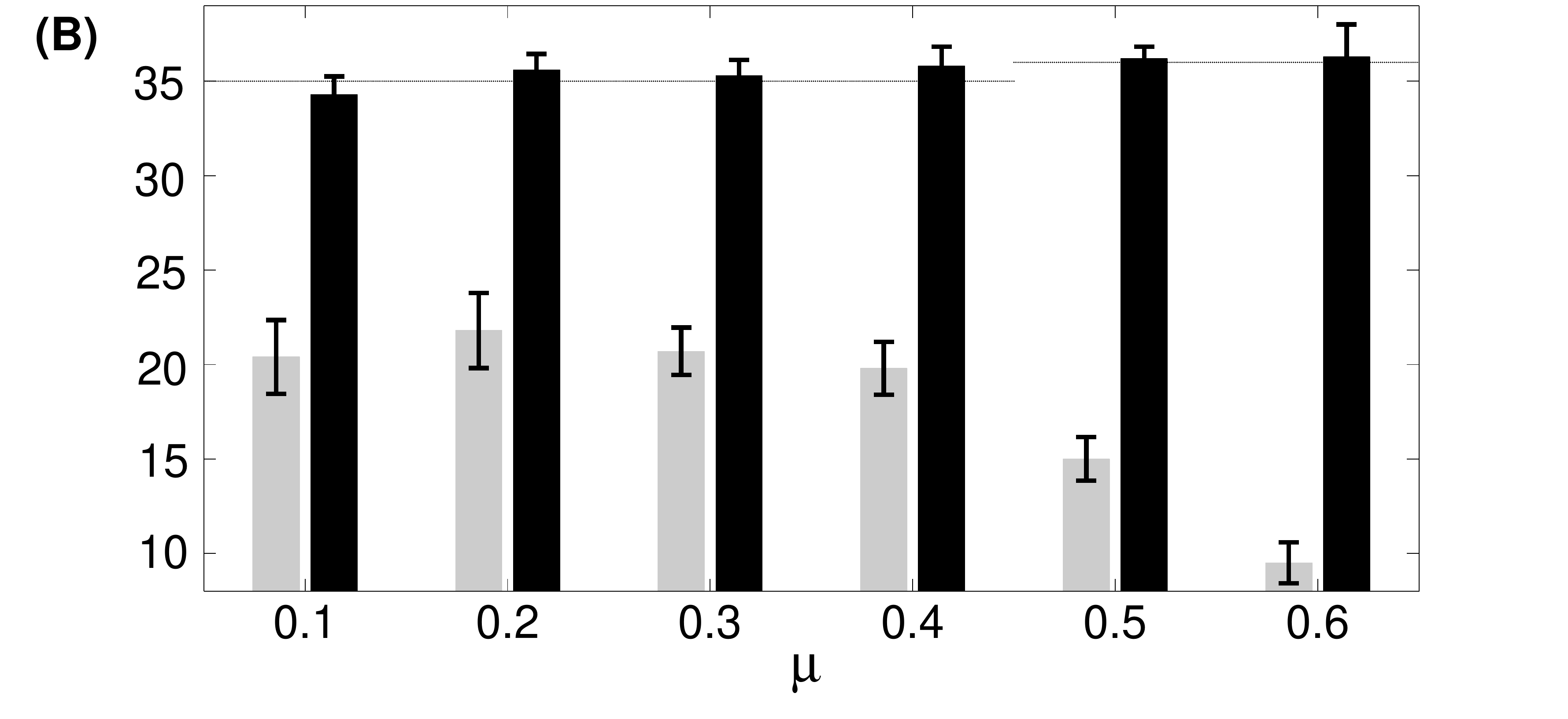}}
\caption{Number of communities estimated by Bayesian NMF and SBMF on GN and non-overlapping LFR networks with the standard deviation errors. The black horizontal line is the actual number of communities.}\label{Fig:09}
\end{figure*}

 In this subsection, we compare the Bayesian NMF \cite{nmf1} with our SBMF model, and give the numerical results of NMI on the GN networks and the non-overlapping LFR networks, and generalized NMI on the overlapping LFR networks. We also compare the abilities of Bayesian NMF and SBMF to infer the appropriate number of communities. The results are averaged over ten trials and are shown in Figs. \ref{Fig:03}$-$\ref{Fig:09}. From these figures, one can observe that: (i) Both the results of the Bayesian NMF and the SBMF model decrease when Zout or $\mu$ are increasing and the standard deviations are low; (ii) SBMF consistently outperforms Bayesian NMF, especially on the non-overlapping LFR networks. For example, when $\mu=0.6$, the NMI of the proposed model (96.32\%) is $35\%$ higher than that of the Bayesian model (60.32\%); (iii) On the overlapping LFR networks, SBMF outperforms the methods in \cite{comparison}, where a systematic comparison of different methods on the overlapping LFR benchmarks with the same parameter settings is conducted. In addition, the generalized NMI do not decrease significantly with the increasing fraction of the overlapping nodes, indicating that our model is applicable for highly overlapping networks; (iv) The inferred community numbers based on the modified partition density are closer to the real ones, and the standard deviations are lower. For example, the community number of the GN networks has been perfectly recovered. Note that the partition density can be applied to both the cases of non-overlapping and overlapping communities, which makes it suitable for a wide range of applications.

\subsection{Experimental Results on Real Networks}
\begin{figure*}
\includegraphics[height=70mm,width=130mm]{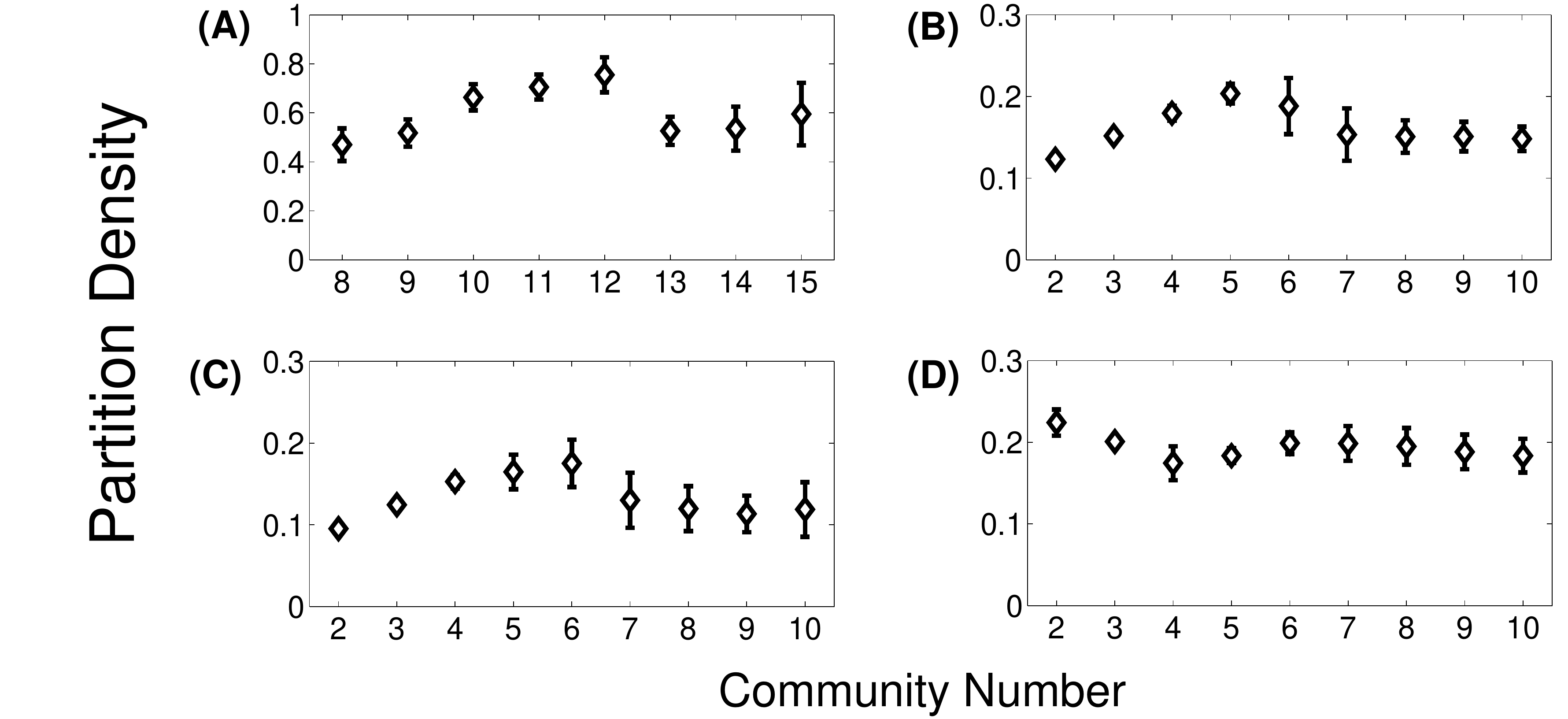}
\caption{Averaged partition density of SBMF versus community number on real world networks: (A) football network; (B) polbooks network; (C) dolphins network; (D) jazz bands network.}\label{Fig:05}
\end{figure*}

In this subsection, we conduct experiments on the real world networks. Figure \ref{Fig:05} gives the results of partition densities under different community numbers $c$, and Table \ref{Tab:04} gives the inferred community numbers of different networks.

\begin{figure*}
\subfigure{\includegraphics[height=50mm,width=75mm]{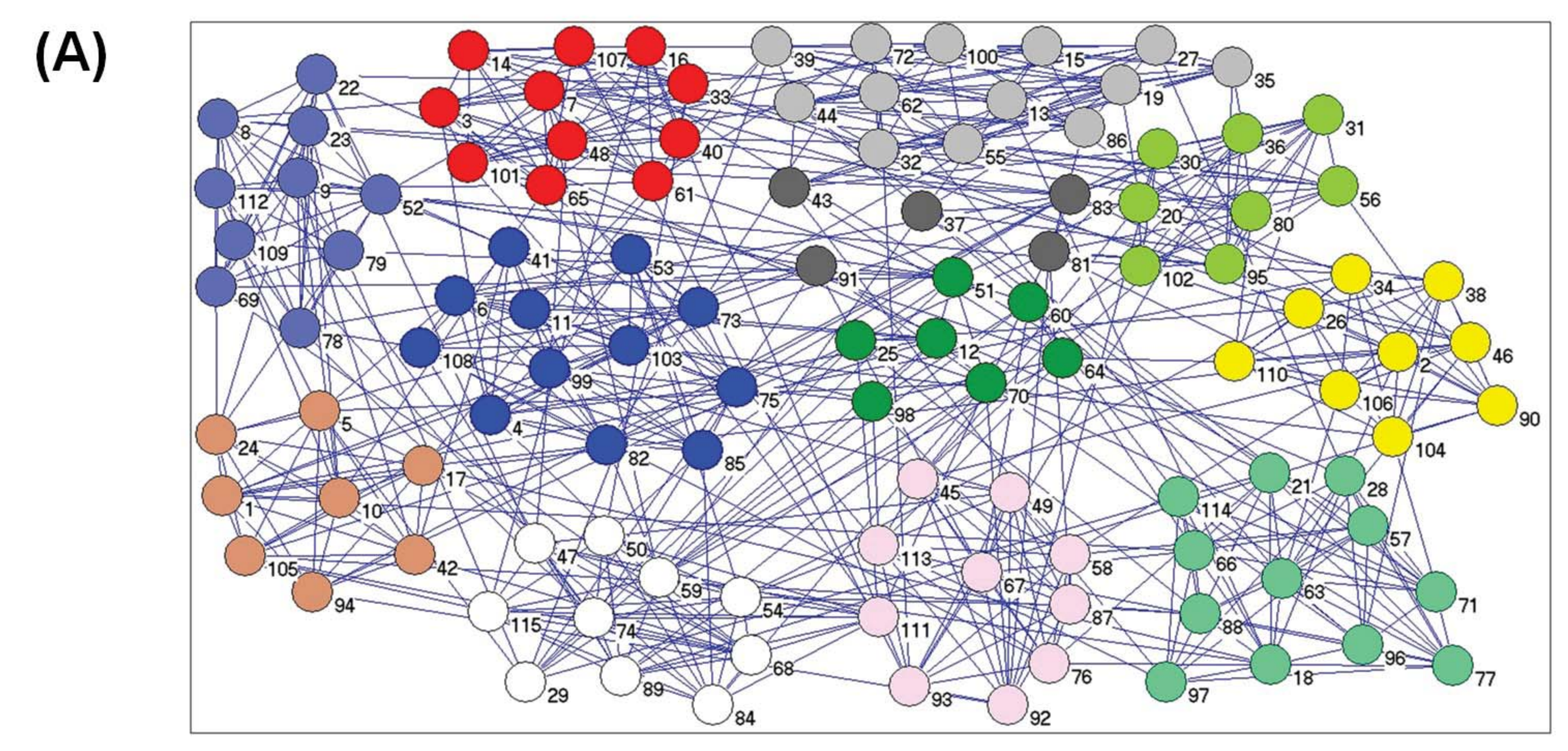}}
\hspace{5mm}
\subfigure{\includegraphics[height=50mm,width=77mm]{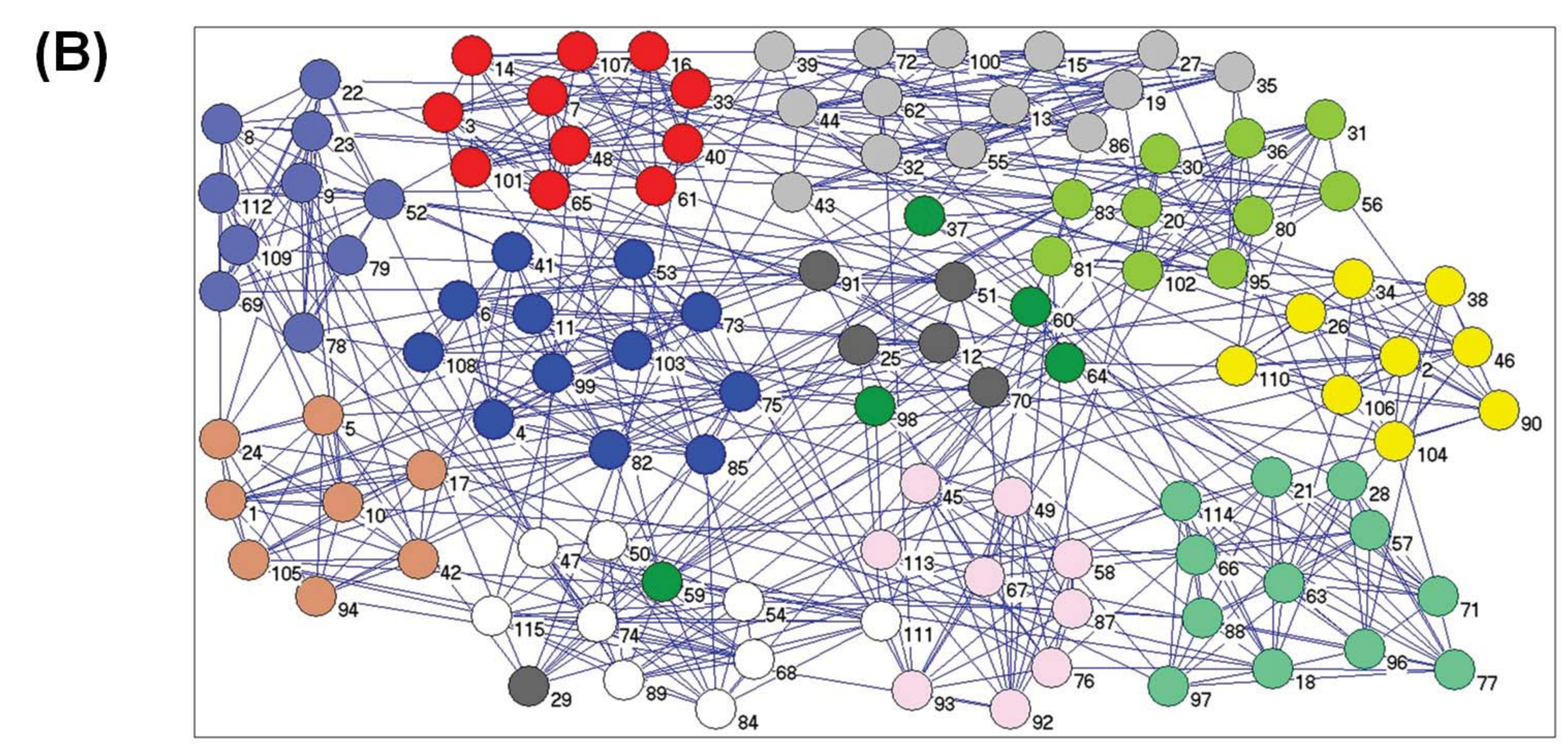}}
\caption{Comparison of (A) real grouping and (B) communities found by SBMF model in the football network. There are no outliers and overlapping teams. All of the mis-clustered teams are abnormal ones, which are listed in Table \ref{Tab:05}.}\label{Fig:07}
\end{figure*}

 We use the football network as a case study. The teams in the network are assigned into 12 conferences, most of which play more games against other ones in the same conference. However, there are some abnormal teams that play
against the ones in other conferences more frequently. Table \ref{Tab:05} lists the
basic information about these teams, among which the teams 37, 43, 81, 83, 91 are in the conference {\it IA Independents} (the black ones in Fig. \ref{Fig:07} (a)), and the teams 12, 25, 51, 60, 64, 70, 98 are in the
conference {\it Sunbelt} (the dark green ones in Fig. \ref{Fig:07} (a), next to the black ones). To determine the best community number $c$, we calculate the partition density given different $c$. The peak value is achieved at $c=12$. The corresponding partitioning result of SBMF is shown in Fig.\ref{Fig:07} (b), from which one can observe that: i) Our proposed BMF model only mis-clusters the abnormal teams; ii) The abnormal teams are reallocated to the other conferences based on the real topology structures; and iii) No outliers or overlapping teams are detected.

\begin{figure}
\hspace{-5mm}
\includegraphics[height=50mm,width=95mm]{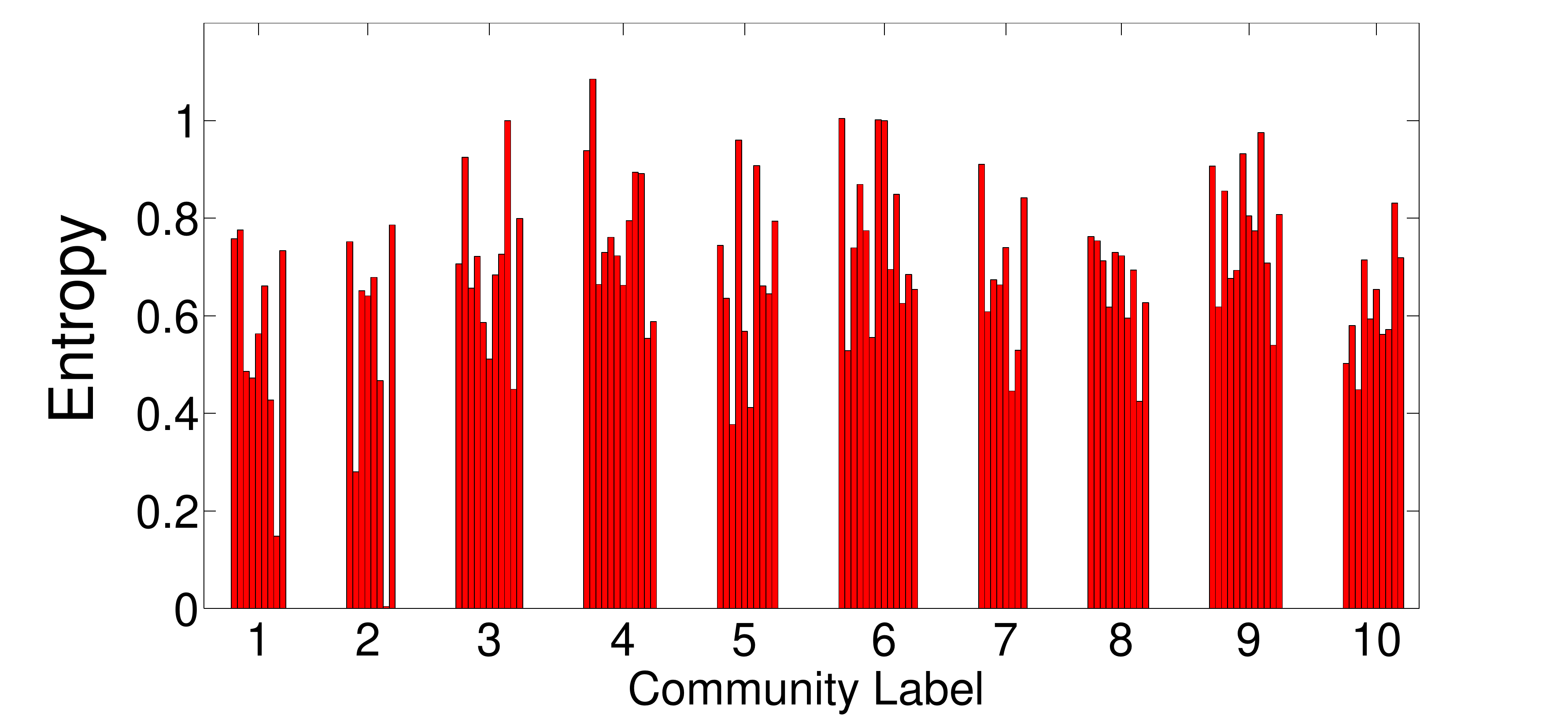}
\caption{Entropy of the nodes in different conferences. The higher the entropy, the fuzzier the membership degrees.}\label{Fig:08}
\end{figure}

Based on the partitioning result, one can analyze the membership degrees of nodes based on the matrix $U$ obtained using Alg. \ref{Al:01}
. For example, one can use the following entropy value to analyze the positions of different nodes in the corresponding communities.
$$
H_i = -\sum_jU_{ij}\log U_{ij}.
$$
The entropies of the nodes in ten normal conferences (i.e., not including the conferences {\it IA Independents} and {\it Sunbelt}) are shown in Fig. \ref{Fig:08}.
The lower the entropy, the more likely the node is to stay in its own community.


\begin{table}[H]
\caption{Number of communities estimated by Bayesian NMF and SBMF on real world networks.}
\centering\scriptsize
\begin{tabular}{c | c c}\hline\hline
 Dataset  & BayesianNMF & SBMF\\\hline
 Football & $9.2\pm 0.63$ & $12\pm0$ \\
 Dolphin & $7.6\pm 0.70$ & $6\pm0$\\
 Polbooks & $6.9\pm 0.57$ & $5\pm0$ \\
 Jazz &  $11.1\pm0.74$ & $2\pm0$ \\\hline\hline
\end{tabular}\label{Tab:04}
\end{table}

\begin{table}[H]
\caption{List of the abnormal teams that played more frequently against the ones in the other conferences. ``N\_S'' and ``N\_O''  mean the times that the team played against the other ones in the same conference and in the other conferences, respectively.}
\centering\scriptsize
\begin{tabular}{c | c c || c | c c}\hline\hline
 Team ID  & \hspace{1mm} N\_S \hspace{1mm} & \hspace{1mm} N\_O \hspace{1mm} & Team ID & \hspace{1mm} N\_S \hspace{1mm} & \hspace{1mm} N\_O \hspace{1mm} \\\hline
 \emph{37} & 0 & 8 & \emph{60} & 2 & 6 \\
 \emph{43} & 0 & 7 & \emph{64} & 2 & 7 \\
 \emph{81} & 1 & 10 & \emph{70} & 3 & 8 \\
 \emph{83} & 1 & 10 & \emph{98} & 3 & 5 \\
 \emph{91} & 0 & 9 & \emph{111} & 0 & 11\\
 \emph{12} & 4 & 6 & \emph{29} & 0 & 9\\
 \emph{25} & 3 & 7 & \emph{59} & 2 & 8\\
 \emph{51} & 3 & 6 &           &   & \\
 \hline\hline
\end{tabular}\label{Tab:05}
\end{table}
    传统的概率不可靠  不利于解释


\section{Conclusions and Future work}\label{conclusion}
In this paper we present a symmetric binary matrix factorization model to detect overlapping community structures. The model can explicitly identify the community memberships of the nodes, which are allowed to belong to multiple communities or to be outliers. We also give a revised partition density to automatically infer the community number in the network. The experiments conducted on both the synthetic and the real world networks showed the effectiveness of the proposed method. In summary, the SBMF model is parameter-free, easy to implement, and the new partition density is good at determining the number of communities.

There are two interesting problems that are worthy of working on in the future: i) How to extend the proposed model to weighted networks and directed networks; and ii) How to combine the model with a recommendation system to get better recommending results.
\begin{acknowledgments}
Z.-Y. Z is supported by the National Natural Science Foundation of China under Grant No. 61203295. Y. W is supported by the National Natural Science Foundation of China under Grant No. 11131009 and 61171007.
\end{acknowledgments}
\bibliography{bipartite}
\end{document}